1

# Genetic, Individual, and Familial Risk Correlates of Brain Network Controllability in Major Depressive Disorder

Tim Hahn [1†], Nils R. Winter [1], Jan Ernsting[1,2], Marius Gruber [1], Marco J. Mauritz [1], Lukas Fisch [1], Ramona Leenings [1,2], Kelvin Sarink [1], Julian Blanke [1], Vincent Holstein [1], Daniel Emden [1], Marie Beisemann [3], Nils Opel [1,4], Dominik Grotegerd [1], Susanne Meinert [1], Walter Heindel [5], Stephanie Witt [6], Marcella Rietschel [6], Markus M. Nöthen[7], Andreas J. Forstner[7,8], Tilo Kircher [9], Igor Nenadic [9], Andreas Jansen [9], Bertram Müller-Myhsok [10], Till F. M. Andlauer [11], Martin Walter [12], Martijn P. van den Heuvel [13,14], Hamidreza Jamalabadi [15], Udo Dannlowski [1*], Jonathan Repple [1*]

[1] Institute for Translational Psychiatry, University of Münster, Germany
[2] Faculty of Mathematics and Computer Science, University of Münster, Germany
[3] Department of Statistics, TU Dortmund University, Dortmund, Germany
[4] Interdisciplinary Centre for Clinical Research IZKF, University of Münster
[5] Institute of Clinical Radiology, University of Münster, Germany
[6] Department of Genetic Epidemiology, Central Institute of Mental Health, Faculty of Medicine Mannheim, University of Heidelberg, Mannheim, Germany
[7] Institute of Human Genetics, University of Bonn, School of Medicine & University Hospital Bonn, Bonn, Germany
[7] Centre for Human Genetics, University of Marburg, Marburg, Germany
[8] Institute of Neuroscience and Medicine (INM-1), Research Center Jülich, Jülich, Germany
[9] Department of Psychiatry and Psychotherapy, Phillips University Marburg, Germany
[10] Max-Planck-Institute of Psychiatry, Munich, Germany
[11] Department of Neurology, Klinikum rechts der Isar, Technical University of Munich, Germany
[12] Department of Psychiatry and Psychotherapy, Jena University Hospital, Jena, Germany
[13] Department of Complex Trait Genetics, Center for Neurogenomics and Cognitive Research, Vrije Universiteit Amsterdam, Amsterdam Neuroscience, Amsterdam, The Netherlands
[14] Department of Child Psychiatry, Amsterdam University Medical Center, Amsterdam Neuroscience, Amsterdam, The Netherlands
[15] Department of Psychiatry and Psychotherapy, University of Tübingen, Tübingen, Germany

* these authors contributed equally to this work

†Corresponding author:

Tim Hahn, Ph.D., Department of Psychiatry, University of Münster, Germany

Albert-Schweitzer-Campus 1, D-48149 Münster

Phone: +49 (0)2 51 / 83 – 56625, Fax: +49 (0)2 51 / 83 – 56612, E-Mail: HahnT@wwu.de

**Abstract**

Background: A therapeutic intervention in psychiatry can be viewed as an attempt to influence the brain's large-scale, dynamic network state transitions underlying cognition and behavior. Building on connectome-based graph analysis and control theory, Network Control Theory is emerging as a powerful tool to quantify network controllability – i.e., the influence of one brain region over others regarding dynamic network state transitions. If and how network controllability is related to mental health remains elusive.

Methods: From Diffusion Tensor Imaging data, we inferred structural connectivity and inferred calculated network controllability parameters to investigate their association with genetic and familial risk in patients diagnosed with major depressive disorder (MDD, n=692) and healthy controls (n=820).

Results: First, we establish that controllability measures differ between healthy controls and MDD patients while not varying with current symptom severity or remission status. Second, we show that controllability in MDD patients is associated with polygenic scores for MDD and psychiatric cross-disorder risk. Finally, we provide evidence that controllability varies with familial risk of MDD and bipolar disorder as well as with body mass index.

Conclusions: We show that network controllability is related to genetic, individual, and familial risk in MDD patients. We discuss how these insights into individual variation of network controllability may inform mechanistic models of treatment response prediction and personalized intervention-design in mental health.

**1. Introduction**

Complex network theory conceptualizes the brain as a dynamical system which depends on the interactions between distributed brain regions (1). Accordingly, the brain can be viewed as an intricate network of brain regions that synchronize their activity via anatomical and functional connections. Based on this, mathematical graph theory is utilized to gain insights into the underlying organizational principles of the brain (2,3) and its topological organization in health and disease (4,5). For example, reduced global fractional anisotropy (FA) has been associated with remission status of depressive patients, while FA in connections between frontal, temporal, insular, and parietal regions was found to be negatively associated with symptom severity (6,7). Cross-disorder connectome analyses have further revealed disruptions in connections central to global network communication and integration, emphasizing the involvement of the connectome in a wide range of mental health and neurological conditions (8). In addition, machine learning on graphs – for example graph convolutional networks or reinforcement learning-based graph dismantling (9) – is emerging as a fruitful extension of classical graph analysis.

While classic connectome analysis has yielded tremendous insights into the topological organization of the brain in health and disease, it does not directly advance our ability to actively manipulate and control the brain. It is, however, this very ability to control the large-scale dynamics of the brain which facilitates virtually all therapeutic interventions in psychiatry. In short, any intervention – from medication to psychotherapy – can be conceptualized as an attempt to control the large-scale, dynamic network state transitions in the brain which underly cognition and behavior (1,10). Indeed, Control Theory as the study and practice of controlling dynamical systems is ubiquitous in medicine and biology (11), framing any intervention – from the optimization of cancer chemotherapy (12,13) and the design of artificial organs (14) to real-time drug administration and non-pharmaceutical pandemic defense strategies (15) – as a control problem.

Relating Control Theory and network neuroscience, recent progress in Network Control Theory has enabled the quantification of the influence a brain region has on the dynamic transitions between brain states underlying cognition and behavior (1,10). This so-called controllability of a brain region is linked to its structural connectivity properties which constrain or support transitions between different brain states (16,17). Controllability of a brain region is commonly captured by two key metrics: On the one hand, *average controllability* quantifies the capacity of a brain region to facilitate transitions to easy-to-reach states such as those emerging at rest. On the other hand, *modal controllability* quantifies the ease with which a brain region can steer the brain into difficult-to-reach states such as those involved in more demanding cognitive tasks.

Elucidating the variation and effect of controllability in mental disorders is of particular interest as controlling large-scale state transitions in the brain that underly cognition and behavior is at the heart of all therapeutic interventions in psychiatry (1,18). Fueled by evidence that the human brain is in principle controllable (16) and the recently discovered associations with cognition (17,19), studies with small to moderate patient sample sizes have begun to investigate network controllability in mental disorders. First, Jeganathan et al. (20) showed altered controllability in young people with bipolar disorder (n=38) and those at high genetic risk (n=84) compared to healthy controls (n=96). Likewise, Braun et al. (21) showed altered network control properties in schizophrenia patients (n=24) as compared to (n=178) healthy controls. Of note, Parkes et al. (22) investigated the association between average controllability and negative and positive psychosis spectrum symptoms in a large sample of youths between 8 and 22 years of age. Related to mental disorders, Kenett et al. (23) showed regional associations between controllability and subclinical depressive symptoms as measured using the Beck Depression Inventory (24) in healthy controls.

Building on these advances, we provide a comprehensive characterization of individual variation in average and modal controllability with regard to demographic, disease-related, genetic, personal, and familial risk in Major Depressive Disorder (MDD). First, we assess the effect of age and gender on average and modal controllability. Then, we compare average and modal controllability between healthy controls and MDD patients and test whether these measures vary with age, gender, current symptom severity, or remission status. Second, we assess whether average and modal controllability in MDD patients are associated with polygenic scores for MDD (25), Bipolar Disorder (26), and psychiatric cross-disorder (27) risk as well as with familial risk of MDD and bipolar disorder. Finally, we quantify the effects of body mass index as a personal risk factors previously reported to be associated with brain-structural deviations in MDD on average and modal controllability (28,29).

## 2. Methods and Materials

Sample

Participants were part of the Marburg-Münster Affective Disorders Cohort Study (MACS) (30) and were recruited at two different sites (Marburg & Münster, Germany). See (31) for a detailed description of the study protocol. Participants ranging in age from 18 to 65 years were recruited through newspaper advertisements and local psychiatric hospitals. All experiments were performed in accordance with the ethical guidelines and regulations and all participants gave written informed consent prior to examination. To confirm the psychiatric diagnosis or a lack thereof, the Structural Clinical Interview for Diagnostic and Statistical Manual of Mental Disorders-IV Text Revision (DSM-IV-TR) (SCID-I; (32)) was used. MDD subjects were included with current acute depressive episodes and partial or full remission from depression. Patients could be undergoing in-patient, out-patient, or no current treatment at all. Exclusion criteria comprised the presence of any neurological abnormalities, history of seizures, head trauma or unconsciousness, severe physical impairment (e.g. cancer, unstable diabetes, epilepsy etc.), pregnancy, hypothyroidism without adequate medication, claustrophobia, color blindness, and general MRI contraindications (e.g. metallic objects in the body). Only Caucasian subjects were included in the analyses. Further, lifetime diagnoses of schizophrenia, schizoaffective disorder, bipolar disorder, or substance dependence posed reason for exclusion. After excluding subjects according to the aforementioned exclusion criteria, DTI data for 1567 subjects were available. 55 subjects were excluded due to poor DTI quality (see below for a detailed description of the quality assurance procedure). Final samples of n=692 MDD patients and n=820 healthy controls were used for the controllability analyses. See Table 1 for a sample description of sociodemographic and clinical data.

Imaging data acquisition

In the MACS Study, two MR scanners were used for data acquisition located at the Departments of Psychiatry at the University of Marburg and the University of Münster with different hardware and software configurations. Both T1 and DTI data were acquired using a 3T whole body MRI scanner (Marburg: Tim Trio, 12-channel head matrix Rx-coil, Siemens, Erlangen, Germany; Münster: Prisma, 20-channel head matrix Rx-coil, Siemens, Erlangen, Germany). A GRAPPA acceleration factor of two was employed. For DTI imaging, fifty-six axial slices, 2.5 mm thick with no gap, were measured with an isotropic voxel size of 2.5 x 2.5 x 2.5 mm³ (TE=90 ms, TR=7300 ms). Five non-DW images (b0=0) and 2 x 30 DW images with a b-value of 1000 sec/mm² were acquired. Imaging pulse sequence parameters were standardized across both sites to the extent permitted by each platform. For a description of MRI quality control procedures see (31). The body coil at the Marburg scanner was replaced during the study. Therefore, a variable modeling three scanner sites (Marburg old body coil, Marburg new body coil and Münster) was used as covariate for all statistical analyses.

Imaging data preprocessing

Connectomes were reconstructed involving the following steps (33). For a more detailed description of the preprocessing see (6). In accordance with (6), we decided on using a basic DTI reconstruction rather than more advanced diffusion direction reconstruction methods to provide a reasonable balance between false negative and false positive fiber reconstructions (34). For each subject an anatomical brain network was reconstructed, consisting of 114 areas of a subdivision of the FreeSurfer's Desikan–Killiany atlas (35,36), and the reconstructed streamlines between these areas. White matter connections were reconstructed using deterministic streamline tractography, based on the Fiber Assignment by Continuous Tracking (FACT) algorithm (37). Network connections were included when two nodes (i.e., brain regions) were connected by at least three tractography streamlines (38). For each participant, the network information was stored in a structural connectivity matrix, with rows and columns reflecting cortical brain regions, and matrix entries representing graph edges. Edges were only described by their presence or absence to create unweighted graphs.

DTI quality control

In accordance with (6), measures for outlier detection included 1. average number of streamlines, 2. average fractional anisotropy, 3. average prevalence of each subject's connections (low value, if the subject has "odd" connections), and 4. average prevalence of each subjects connected brain regions (high value, if the subject misses commonly found connections). For each metric the quartiles (Q1, Q2, Q3) and the interquartile range (IQR=Q3-Q1) was computed across the group and a datapoint was declared as an outlier if its value was below Q1-1.5*IQR or above Q3+1.5*IQR on any of the four metrics.

Genotyping and calculation of polygenic scores

Genotyping was conducted using the PsychArray BeadChip (Illumina, San Diego, CA, USA), followed by quality control and imputation, as described previously (39,40). In brief, quality control and population substructure analyses were performed in PLINK v1.90 (41), as described in the Supplementary Methods. The data were imputed to the 1000 Genomes phase 3 reference panel using SHAPEIT and IMPUTE2.

For the calculation of polygenic risk scores (PRS; (42)), single-nucleotide polymorphism (SNP) weights were estimated using the PRS-CS method (43) with default parameters. This method employs Bayesian regression to infer PRS weights while modeling the local linkage disequilibrium patterns of all SNPs using the EUR super-population of the 1000 Genomes reference panel. The global shrinkage parameter ϕ was determined automatically (PRS-CS-auto; CD: ϕ=1.80×10$^{-4}$, MDD: ϕ=1.11×10$^{-4}$). The PRS were calculated, using these weights, in PLINK v1.90 on imputed dosage data based on summary statistics of genome-wide association studies (GWAS) by the Psychiatric Genomics Consortium (PGC) containing 162,151 cases and 276,846 controls for a cross-disorder phenotype (27) and 59,851 cases and 113,154 controls for MDD (25). PRS were available for 637 of the 692 MDD patients.

Network Controllability Analysis

To assess the ability of a certain brain region to influence other regions in different ways, we adopt the control theoretic notion of controllability. Controllability of a dynamical system refers to the possibility of driving the state of a dynamical system to a specific target state by means of an external control input (44). A state is defined as the vector of neurophysiological activity magnitudes across brain regions at a single time point. In this paper, following the established model of structural brain controllability (16), we assume the system to follow a noise-free linear time-invariant model as in equation 1.

$$x(k+1) = Ax(k) + Bu(k) \quad (1)$$

where $x$ represents the temporal activity of the 114 brain regions, $A$ is the adjacency matrix whose elements quantify the structural connectivity between every two brain regions, $B$ is the input matrix and $u$ shows the control strategy. Classic results in control theory ensure that controllability of the network is equivalent to the controllability Gramian matrix $W_j = \sum_{i=0}^{\infty} A^i BB^T (A^T)^i$ where $T$ denotes matrix transpose. A rigorous mathematical formulation of network controllability in brain networks can be found in (16). From the Gramian matrix, different controllability measures can be computed for each node (brain region) in the network. Here, based on previous research of network controllability in brain networks, we compute for each participant and each brain region their average controllability and modal controllability as defined in (16).

Average Controllability identifies brain regions that, on average, can drive the system into different states with low input energy and is estimated as the trace of the controllability Gramian matrix i.e. $Tr(W_j)$. Thus, regions with high average controllability can move the brain to easily reachable states. Previous work has identified brain regions that demonstrate high average controllability, such as the precuneus, posterior cingulate, superior frontal, paracentral, precentral, and subcortical structures (16).

Modal Controllability (MC) identifies brain regions that can drive the brain into different states that require high input energy to achieve and is estimated as $\phi_j = \sum_n^N [1 - \xi_n^2(A)] v_{nj}^2$ where $\xi_j$ and $v_{nj}$ represent respectively the eigenvalues and elements of the eigenvector matrix of $A$ ($N = 114$). Thus, regions with high modal controllability can move the brain to difficult to reach states. Previous work has identified brain regions that demonstrate high modal controllability, such as the postcentral, supramarginal, inferior parietal, pars orbitalis, medial orbitofrontal, and rostral middle frontal cortices (16).

Building on these definitions, we estimate single node controllability measures (average and modal controllability) by setting $B = e_j$ where $e_j$ is the $j^{th}$ canonical vector. Whole-brain controllability is then defined as the average of single node controllability measures over all nodes.

Statistical Analyses

Our analysis process is as follows (Figure 1): Based on DTI data (Figure 1a), we defined anatomical brain networks by subdividing the entire brain into 114 anatomically distinct brain regions (network nodes) in a commonly used anatomical atlas (35,36). Following prior work (see *Imaging Data Preprocessing*), we connected nodes (brain regions) by the number of white matter streamlines which results in sparse, undirected structural brain networks for each participant (Figure 1b). Next, a simplified model of brain dynamics was applied to simulate network control and quantify average and modal controllability for each brain region for each participant, as described in (16,44). Figure 1c illustrates the dynamic state transitions of the brain over time. Note that in our analyses, a brain state is characterized not by three, but 114 values per time point, corresponding to the 114 regions contained in the atlas. Points in this space (colored points in Figure 1c) thus correspond to brain states at different time points. Controllability parameters are, in turn, related to the ease with which a given brain region can induce dynamic state transitions in this space (see *Network Controllability Analysis*).

We then analyzed mean whole-brain controllability as well as regional (i.e., per-node) controllability (dependent variable) using an ANCOVA approach with age, gender, MRI scanner site, and the number of present edges as covariates. For all analyses involving PRS, we also controlled for ancestry (first three MDS components). Also, we removed outliers defined as values located more than three standard deviations from the mean. In all analyses involving MDD patients only, we additionally controlled for medication load in accordance with previous publications (6,45). Note that, in the regional analyses, we controlled for multiple comparisons by calculating the false discovery rate (46) with a false-positive rate of .05. All other constructs (age, gender, diagnosis, symptom severity, familial risk of MDD and BD as well as polygenic scores for MDD, BD, and cross-disorder risk) were tested independently and not corrected further for multiple testing.

**3. Results**

Demographic Effects

First, we examined whether chronological age and gender are associated with controllability as has been shown before (19,47). Indeed, we find that the *whole-brain average controllability* was negatively correlated with age for both healthy controls (F(1,811)=24.47, p<.001) and MDD patients (F(1,686)=15.08, p<.001). Likewise, the *regional average controllability* significantly varied with age in 30 and 35 different regions for healthy controls and MDD patients, respectively (all p<.05, FDR-corrected; for a full list of regions for all analyses yielding significant regional associations, see Supplementary Results Tables S1 to S16). *Whole-brain modal controllability* was positively correlated with age for both healthy controls (F(1,811)=3.93, p=.048) and showed a similar trend in MDD patients (F(1,685)=2.91, p=.089). *Regional modal controllability* significantly varied with age in 33 regions for both healthy controls and MDD patients (all p<.05, FDR-corrected). Gender was not significantly associated with *whole-brain average controllability* for healthy controls (F(1,814)=.04, p=.839) or MDD patients (F(1,687)=.08, p=.773). In contrast, *regional average controllability* significantly varied with gender in 12 and 13 regions for healthy controls and MDD patients, respectively (all p<.05, FDR corrected). *Whole-brain modal controllability* was higher in males than in females for healthy controls (F(1,814)=7.58, p=.006) and showed a similar trend in MDD patients (F(1,687)=3.73, p=.054). *Regional modal controllability* significantly varied with age in 16 and 18 regions for healthy controls and MDD patients, respectively (all p<.05, FDR corrected).

Disease-related Variation

Focusing specifically on controllability in MDD, we show that patients displayed lower whole-brain *modal controllability* (F(1,1505)=7.96, p=.005) than healthy controls. Correspondingly, we observed a non-significant trend towards higher whole-brain *average controllability* values in MDD patients (F(1,1505)=3.08, p=.080).

In contrast to previous findings in sub-clinically depressed controls (23), our results do not support an effect of current symptom severity, as measured by the Beck Depression Inventory, on the whole-brain *average* (F(1,671)=.19, p=.665) or *modal* controllability (F(1,671)=1.49, p=.222) in MDD patients. In line with this observation, the remission status was neither associated with *average* (F(2,683)=.43, p=.649) nor *modal controllability* (F(2,683)=.07, p=.935) on the whole-brain or regional level in MDD patients. For direct comparison with the previous publication, we also analyzed the healthy controls only: Again, we did not find a significant association between current symptom severity and whole-brain *average* (F(1,792)=.89, p=.347) or *modal controllability* (F(1,792)=2.36, p=.125).

Genetic and Familial Risk Factors

Next, we examined whether controllability in MDD patients is associated with familial risk of MDD and Bipolar Disorder. We show that *average controllability* was significantly higher in patients carrying self-reported familial risk of MDD (F(1,685)=4.87, p=.028), mirroring the trend-wise increased average controllability of MDD patients compared to healthy controls. This was not the case for *modal controllability* (F(1,685)=2.40, p=.122).

*Average controllability* was also higher in patients carrying a familial risk of Bipolar Disorder (F(1,685)=10.30, p=.001) with regional effects in the right supramarginal gyrus, right inferior parietal gyrus, and precuneus. Likewise, whole-brain *modal controllability* was lower in patients carrying a familial risk of Bipolar Disorder (F(1,685)=9.69, p=.002) with regional effects in the right supramarginal gyrus (for detailed regional analyses, see the Supplementary Tables S16 and S17).

Building on this evidence, we extended the analysis to polygenic risk scores and show that polygenic risk scores for MDD (25) were positively correlated with average *controllability* (F(1,623)=3.86, p=.050) as well as negatively associated with *modal controllability* (F(1,624)=4.88, p=.028). Likewise, polygenic risk scores for cross-disorder risk (27) were negatively associated with whole-brain *modal* (F(1,623)=4.17, p=.042), but not *average controllability* (F(1,622)=.99, p=.320). In contrast to the observed effect for familial risk of Bipolar Disorder, we neither found a significant association of average (F(1,622)=.54, p=.462) nor modal controllability (F(1,621)=.11, p=.742) with polygenic risk score for Bipolar Disorder (26).

Body Mass Index

With mounting evidence pointing towards brain-structural deviations relating body mass index and MDD (28,29), we examined the effects of body mass index on controllability. For *average controllability*, we found associations in 9 regions (p<.05, FDR corrected) including negative correlations in the left superior frontal and posterior cingulate gyrus as well as positive correlations in the superior temporal and left lingual gyrus (see Supplementary Tables S13 und S14). With positive and negative regional associations, a whole-brain effect was absent (F(1,643)=.31, p=.579). Analyses of *modal controllability* revealed the involvement of 6 regions (p<.05, FDR-corrected) showing a similar set of regions including the left superior frontal, posterior cingulate, and superior temporal gyrus with – as expected – a switched direction of correlations and, again, no consistent whole-brain effect (F(1,643)=1.69, p=.194).

## 3. Discussion

Building on Network Control Theory, we investigated the association of average and modal network controllability with genetic, familial, and individual risk in MDD patients (n=692) and healthy controls (n=820). First, we established that controllability measures differ between healthy controls and MDD patients while not varying with current symptom severity or remission status. Second, we showed that modal and average controllability in MDD patients could be predicted based on polygenic scores for MDD and psychiatric cross-disorder risk as well as associations with familial risk of MDD and bipolar disorder. Finally, we provide evidence that controllability varies with body mass index. This evidence suggests that individual differences in these variables either impact the brain's control architecture (e.g., in the case of genetic effects) or are driven by it – as may be the case for e.g. body mass index.

Against this background, our results indicate that individual differences in demographic, disease-related, genetic, individual, and familial risk factors are associated with controllability. We replicated previous findings showing that age and gender affected controllability measures (19,47) also for MDD patients. Given that women are disproportionally affected by MDD, future studies might investigate gender differences in more detail.

Interestingly, associations were mainly found with whole-brain controllability – modal and average alike – suggesting subtle changes in how effectively not only single regions, but a larger set of regions in the brain can drive state transitions. This is of particular interest as previous studies have focused on the set of 30 regions with the highest controllability defined a priori, thereby potentially obscuring such whole-brain effects. This suggests that extending current controllability analyses towards the investigation of sets of regions controlling the brain (as has been done by, e.g., (48)) might be fruitful also for MDD. Moreover, all results were corrected for the number of present edges, which suggests a specific control effect that goes beyond basic graph properties.

From the more general perspective of control, answering what changes in the brain after a specified stimulation event and which regions are most effective or efficient to stimulate is crucial for all therapeutic interventions. First attempts to predict stimulation outcome in the context of electrical brain stimulation have recently been successful (49). In this context, our results imply that individual characteristics may be relevant when designing future interventions based on Network Control Theory. In turn, our results suggest that variation in response to treatment – e.g., with transcranial magnetic stimulation or electroconvulsive therapy – might be explained by controllability differences arising from demographic, disease-related, genetic, personal, and familial risk. Future studies may therefore investigate whether interventions guided by Network Control Theory are more effective or efficient than current approaches.

Several limitations should be noted. First, calculation of average and modal controllability relies on the simplified noise-free linear discrete-time and time-invariant network model employed in virtually all work on brain Network Control Theory (16,18,50). Given the brain's clearly non-linear dynamics, this is justified as 1) nonlinear behavior may be accurately approximated by linear behavior (51) and 2) the controllability of linear and nonlinear systems is related such that a controllable linearized system is locally controllable in the nonlinear case (see also (16) for details).

Second, our estimation of controllability is based upon Diffusion Tensor Imaging (DTI) tractography which in itself is limited in its ability to accurately quantify the structural connectome (for an introduction, see (52)). Currently, several novel approaches to controllability quantification are being explored including estimation from gray matter (53) and resting-state functional dynamics (50). Empirically comparing and theoretically reconciling results from these methods will be crucial for robust parameter estimation in Network Control Theory studies of the brain. In addition, longitudinal data from DTI, gray matter, and resting-state functional dynamics available from, e.g., the Marburg-Münster Affective Disorders Cohort Study (MACS; (31)) will enable us to assess the (differential) reliability of these approaches. In combination with functional Magnetic Resonance Imaging, this approach also provides an opportunity to further characterize the relationship between network control and individual task-related activation (54).

Third, it should be noted that most effect sizes observed in this study were small. Methodologically, however, it has been shown that small samples systematically inflate the apparent effect size, whereas large samples such as this one provide a much more accurate estimate of the true effect size (55). Most importantly, our characterization of individual differences in controllability in MDD does not consider isolated effects but is supported by a broad range of analyses.

In summary, we build on a growing body of literature studying cognition and psychopathology within the framework of Network Control Theory to show effects of demographic, disease-related, genetic, personal, and familial risk on modal and average controllability in MDD patients. Thereby, we hope to aid future studies employing Network Control Theory to predict treatment response, guide therapeutic planning, and design novel interventions for MDD.

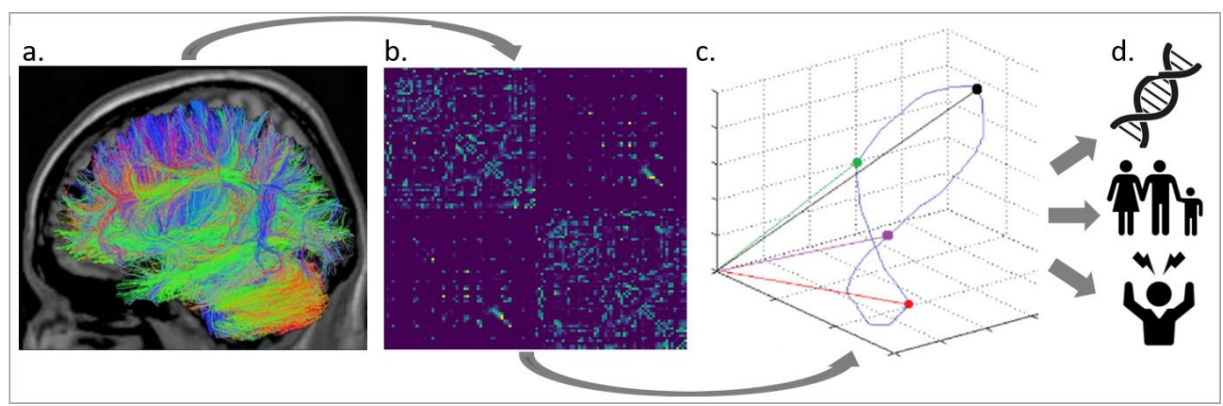

Figure 1. From Diffusion Tensor Imaging data (a), we derived the structural connectivity matrix for each participant (b) and quantified modal and average controllability – i.e., the influence a brain region has on the dynamic transitions between brain states underlying cognition and behavior (c). We then investigated their association with genetic, familial, and personal risk (d).

| Characteristic | MDD[a] (*n* = 692) | HC[a] (*n* = 820) | *p* |
|---|---|---|---|
| Sociodemographic | | | |
| Gender | 451 female, 241 male | 529 female, 291 male | <.001[b] |
| Age, years | 36.41 ± 13.13 | 33.96 ± 12.75 | <.001[c] |
| Questionnaires | | | |
| BDI | 17.67 ± 11.06 | 4.02 ± 4.18 | <.001[c] |
| Clinical | | | |
| Depressive episodes | 3.86 ± 6.28 | – | – |
| Duration of illness, years | 10.27 ±9.75 | – | – |
| Medication | | | |
| Medication load | 1.32 ± 1.47 | – | – |
| CPZ | 24.77 ± 89.21 | – | – |

Table 1. Sample summary. [a]Numbers present either absolute numbers or mean plus standard deviation, [b]$\chi^2$-test (two-tailed), [c]t-test (two-tailed), HC = healthy control group, MDD = patient group with major depression disorder, BDI = sum score based on 21 items, CPZ = chlorpromazine-equivalent doses

Funding

This work was funded by the German Research Foundation (DFG grants HA7070/2-2, HA7070/3, HA7070/4 to TH) and the Interdisciplinary Center for Clinical Research (IZKF) of the medical faculty of Münster (grants Dan3/012/17 to UD, SEED 11/19 to NO, and MzH 3/020/20 to TH). H.J. was supported by body jgrant of Medical Faculty of University of Tübingen (No. 2487-1-0).

The MACS dataset used in this work is part of the German multicenter consortium "Neurobiology of Affective Disorders. A translational perspective on brain structure and function", funded by the German Research Foundation (Deutsche Forschungsgemeinschaft DFG; Forschungsgruppe/Research Unit FOR2107). Principal investigators (PIs) with respective areas of responsibility in the FOR2107 consortium are: Work Package WP1, FOR2107/MACS cohort and brainimaging: Tilo Kircher (speaker FOR2107; DFG grant numbers KI 588/14-1, KI 588/14-2), Udo Dannlowski (co-speaker FOR2107; DA 1151/5-1, DA 1151/5-2), Axel Krug (KR 3822/5-1, KR 3822/7-2), Igor Nenadic (NE 2254/1-2), Carsten Konrad (KO 4291/3-1). WP2, animal phenotyping: Markus Wöhr (WO 1732/4-1, WO 1732/4-2), Rainer Schwarting (SCHW 559/14-1, SCHW 559/14-2). WP3, miRNA: Gerhard Schratt (SCHR 1136/3-1, 1136/3-2). WP4, immunology, mitochondriae: Judith Alferink (AL 1145/5-2), Carsten Culmsee (CU 43/9-1, CU 43/9-2), Holger Garn (GA 545/5-1, GA 545/7-2). WP5, genetics: Marcella Rietschel (RI 908/11-1, RI 908/11-2), Markus Nöthen (NO 246/10-1, NO 246/10-2), Stephanie Witt (WI 3439/3-1, WI 3439/3-2). WP6, multi method data analytics: Andreas Jansen (JA 1890/7-1, JA 1890/7-2), Tim Hahn (HA 7070/2-2), Bertram Müller-Myhsok (MU1315/8-2), Astrid Dempfle (DE 1614/3-1, DE 1614/3-2). CP1, biobank: Petra Pfefferle (PF 784/1-1, PF 784/1-2), Harald Renz (RE 737/20-1, 737/20-2). CP2, administration. Tilo Kircher (KI 588/15-1, KI 588/17-1), Udo Dannlowski (DA 1151/6-1), Carsten Konrad (KO 4291/4-1). Data access and responsibility: All PIs take responsibility for the integrity of the respective study data and their components. All authors and coauthors had full access to all study data. The FOR2107 cohort project (WP1) was approved by the Ethics Committees of the Medical Faculties, University of Marburg (AZ: 07/14) and University of Münster (AZ: 2014-422-b-S).

Supplementary Material for

Genetic, Personal, and Familial Risk Correlates of Brain Network Controllability in Major Depressive Disorder

**Key words:** Network Control Theory; diffusion magnetic resonance imaging, structural connectivity; Major Depressive Diosorder

## Supplementary Methods

**Supplementary M1. Genotyping, quality control, and imputation.**

Genotyping was conducted using the Infinium PsychArray BeadChip, as described previously.[1] The quality control (QC) of genetic data was conducted in PLINK v1.90b6.10[2] and R v3.5.2, as described previously.[3] Pre-imputation QC of genotype data consisted of the following steps:

1. Removal of SNPs with call rates <98% or a minor allele frequency (MAF) <1%
2. Removal of individuals with genotyping rates <98%
3. Removal of sex mismatches
4. Removal of genetic duplicates
5. Removal of cryptic relatives with pi-hat≥12.5
6. Removal of genetic outliers with a distance from the mean of >4 SD in the first eight multidimensional scaling (MDS) ancestry components
7. Removal of individuals with a deviation of the autosomal or X-chromosomal heterozygosity from the mean >4 SD
8. Removal of non-autosomal variants
9. Removal of SNPs with call rates <98% or a MAF <1% or Hardy-Weinberg Equilibrium (HWE) test $p$-values $<1\times10^{-6}$
10. Removal of A/T and G/C SNPs
11. Update of variant IDs and positions to the IDs and positions in the 1000 Genomes Phase 3 reference panel
12. Alignment of alleles to the reference panel
13. Removal of duplicated variants and variants not present in the reference panel

For the calculation of ancestry components (used to determine genetic outliers and as covariates in the analyses), pre-imputation genotype data were used. Additional variant filtering steps were: removal of variants with a MAF <0.05 or HWE $p$-value $<10^{-3}$; removal of variants mapping to the extended MHC region (chromosome 6, 25-35 Mbp) or to a typical inversion site on chromosome 8 (7-13 Mbp); linkage disequilibrium (LD) pruning (command --indep-pairwise 200 100 0.2). Next, the pairwise identity-by-state (IBS) matrix of all individuals was calculated using the command --genome on the filtered genotype data. Multidimensional scaling (MDS) analysis was performed on the IBS matrix using the eigendecomposition-based algorithm in PLINK v1.90b6.10.

After imputation, variants with a MAF <1%, an HWE test p<1×10-6, and an INFO metric <0.8 were removed. Imputation was conducted using SHAPEIT v2 (r837)[4], IMPUTE2 v2.3.2[5,6], and the 1000 Genomes Phase 3 reference panel.

In total, imputed genetic data were available for 2,248 individuals.

Variants before QC: 596,861; variants after QC: 284,691; variants after imputation: 8,565,143.

**Supplementary M2. Calculation of polygenic scores.**

Two PGSs were calculated using training summary statistics from published genome-wide association studies (GWASs): psychiatric cross-disorder (CD)[7] with 162,151 cases and 276,846 controls and MDD[8] (without 23andMe) with 59,851 cases and 113,154 controls.

PGSs were calculated using the PRS-CS[9] method that employs Bayesian regression to infer PGS weights while modeling local linkage disequilibrium (LD) patterns using the 1000 Genomes EUR reference panel. All training GWAS variants with an INFO metric <0.6, a MAF <1%, or which were not present in the FOR2107 imputation were removed before estimating PRS-CS weights. The global shrinkage parameter was determined using the automatic method. PGSs were calculated in *R* using the PRS-CS weights and imputed dosage data, as described previously[10].

# Supplementary Results

Supplementary Table S1. Regional average controllability association with chronological age in healthy controls.

| DV | F | p_cor | p |
|---|---|---|---|
| ctx_rh_paracentral_1 | 39.913204 | 5.010844e-08 | 4.395477e-10 |
| ctx_lh_caudalanteriorcingulate_1 | 27.506280 | 1.141134e-05 | 2.001989e-07 |
| ctx_lh_paracentral_1 | 23.140227 | 5.320771e-05 | 1.796351e-06 |
| ctx_rh_caudalanteriorcingulate_1 | 23.065264 | 5.320771e-05 | 1.866937e-06 |
| ctx_rh_posteriorcingulate_1 | 20.449947 | 1.604855e-04 | 7.038838e-06 |
| ctx_lh_posteriorcingulate_1 | 18.830596 | 3.058329e-04 | 1.609647e-05 |
| ctx_rh_superiortemporal_1 | 18.195250 | 3.632203e-04 | 2.230300e-05 |
| ctx_lh_rostralmiddlefrontal_3 | 17.094047 | 5.603898e-04 | 3.932560e-05 |
| ctx_lh_precentral_1 | 16.573607 | 6.511334e-04 | 5.140527e-05 |
| ctx_lh_superiorfrontal_1 | 16.306559 | 6.732191e-04 | 5.905430e-05 |
| ctx_rh_inferiorparietal_2 | 13.746358 | 2.316970e-03 | 2.235673e-04 |
| ctx_lh_superiorparietal_3 | 12.943547 | 3.235487e-03 | 3.405775e-04 |
| ctx_lh_parstriangularis_1 | 11.793686 | 5.276616e-03 | 6.247479e-04 |
| ctx_rh_frontalpole_1 | 11.725695 | 5.276616e-03 | 6.480054e-04 |
| ctx_rh_inferiortemporal_2 | 11.503259 | 5.540534e-03 | 7.290176e-04 |
| ctx_lh_frontalpole_1 | 10.564717 | 8.554421e-03 | 1.200621e-03 |
| ctx_rh_precentral_3 | 10.444146 | 8.587628e-03 | 1.280611e-03 |
| ctx_rh_inferiortemporal_1 | 10.293698 | 8.790763e-03 | 1.388015e-03 |
| ctx_rh_pericalcarine_1 | 9.994235 | 9.429103e-03 | 1.629049e-03 |
| ctx_lh_inferiorparietal_2 | 9.965975 | 9.429103e-03 | 1.654229e-03 |
| ctx_lh_inferiortemporal_1 | 9.369004 | 1.238049e-02 | 2.280617e-03 |
| ctx_rh_cuneus_1 | 9.202388 | 1.292634e-02 | 2.494557e-03 |
| ctx_lh_supramarginal_2 | 8.567306 | 1.744102e-02 | 3.518802e-03 |
| ctx_lh_pericalcarine_1 | 7.976473 | 2.306824e-02 | 4.856472e-03 |
| ctx_rh_precuneus_1 | 7.708478 | 2.564330e-02 | 5.623531e-03 |
| ctx_rh_parstriangularis_1 | 7.622326 | 2.585760e-02 | 5.897348e-03 |
| ctx_rh_medialorbitofrontal_1 | 7.527569 | 2.623145e-02 | 6.212711e-03 |
| ctx_lh_parsorbitalis_1 | 6.972400 | 3.436050e-02 | 8.439422e-03 |
| ctx_rh_superiorparietal_2 | 6.437731 | 4.465389e-02 | 1.135932e-02 |
| ctx_rh_precuneus_2 | 6.308580 | 4.639505e-02 | 1.220922e-02 |

Supplementary Table S2. Regional average controllability association with chronological age in MDD patients.

| DV | F | p_cor | p |
|---|---|---|---|
| ctx_lh_paracentral_1 | 27.615740 | 0.000022 | 1.981889e-07 |
| ctx_rh_caudalanteriorcingulate_1 | 26.259969 | 0.000022 | 3.886961e-07 |
| ctx_rh_inferiorparietal_2 | 20.888907 | 0.000220 | 5.785238e-06 |
| ctx_rh_paracentral_1 | 20.144597 | 0.000240 | 8.425743e-06 |
| ctx_rh_fusiform_1 | 19.694432 | 0.000242 | 1.060566e-05 |
| ctx_rh_medialorbitofrontal_1 | 18.431959 | 0.000383 | 2.017327e-05 |
| ctx_lh_superiorfrontal_1 | 16.180397 | 0.000973 | 6.407324e-05 |
| ctx_rh_rostralmiddlefrontal_2 | 16.057289 | 0.000973 | 6.826222e-05 |
| ctx_lh_supramarginal_1 | 15.455737 | 0.001180 | 9.319076e-05 |
| ctx_rh_superiortemporal_1 | 14.206094 | 0.002030 | 1.781118e-04 |
| ctx_lh_caudalanteriorcingulate_1 | 13.204122 | 0.003113 | 3.003896e-04 |
| ctx_lh_precentral_4 | 12.941279 | 0.003274 | 3.446176e-04 |
| ctx_lh_rostralmiddlefrontal_3 | 11.634278 | 0.006015 | 6.859497e-04 |
| ctx_rh_superiorparietal_2 | 10.664269 | 0.009338 | 1.146808e-03 |
| ctx_lh_precentral_1 | 9.760702 | 0.014123 | 1.858258e-03 |
| ctx_lh_posteriorcingulate_1 | 9.482261 | 0.015115 | 2.158486e-03 |
| ctx_lh_lingual_2 | 9.402110 | 0.015115 | 2.253924e-03 |
| ctx_rh_lingual_1 | 9.272367 | 0.015305 | 2.416514e-03 |
| ctx_rh_superiorfrontal_1 | 8.896460 | 0.017759 | 2.959795e-03 |
| ctx_rh_lingual_2 | 8.605041 | 0.019761 | 3.466860e-03 |
| ctx_rh_frontalpole_1 | 8.492462 | 0.019998 | 3.683883e-03 |
| ctx_lh_precuneus_2 | 8.230402 | 0.021050 | 4.246826e-03 |
| ctx_rh_superiorparietal_3 | 8.220011 | 0.021050 | 4.270915e-03 |
| ctx_rh_isthmuscingulate_1 | 8.152536 | 0.021050 | 4.431553e-03 |
| ctx_rh_pericalcarine_1 | 8.069832 | 0.021138 | 4.635479e-03 |
| ctx_rh_parahippocampal_1 | 7.711206 | 0.024410 | 5.639479e-03 |
| ctx_lh_lateraloccipital_1 | 7.665866 | 0.024410 | 5.781294e-03 |
| ctx_lh_rostralanteriorcingulate_1 | 7.315190 | 0.028535 | 7.008539e-03 |
| ctx_lh_superiortemporal_2 | 6.996749 | 0.031218 | 8.353285e-03 |
| ctx_rh_rostralanteriorcingulate_1 | 6.985325 | 0.031218 | 8.407886e-03 |
| ctx_rh_inferiortemporal_1 | 6.966639 | 0.031218 | 8.496357e-03 |
| ctx_rh_parstriangularis_1 | 6.910711 | 0.031218 | 8.763073e-03 |
| ctx_lh_inferiortemporal_1 | 6.656228 | 0.034855 | 1.008966e-02 |
| ctx_lh_cuneus_1 | 6.243881 | 0.042235 | 1.269819e-02 |
| ctx_rh_inferiorparietal_3 | 6.206499 | 0.042235 | 1.296693e-02 |

Supplementary Table S3. Regional modal controllability association with chronological age in healthy controls.

| DV | F | p_cor | p |
|---|---|---|---|
| ctx_rh_paracentral_1 | 40.991466 | 2.948105e-08 | 2.586057e-10 |
| ctx_rh_superiortemporal_1 | 25.346231 | 3.366946e-05 | 5.906923e-07 |
| ctx_rh_inferiorparietal_2 | 22.166978 | 1.117028e-04 | 2.939548e-06 |
| ctx_rh_caudalanteriorcingulate_1 | 21.356275 | 1.263540e-04 | 4.433475e-06 |
| ctx_lh_rostralmiddlefrontal_3 | 20.314469 | 1.719290e-04 | 7.540747e-06 |
| ctx_lh_superiorfrontal_1 | 15.773565 | 1.476721e-03 | 7.772217e-05 |
| ctx_lh_paracentral_1 | 15.120055 | 1.777431e-03 | 1.091405e-04 |
| ctx_rh_precuneus_2 | 14.067286 | 2.499378e-03 | 1.889465e-04 |
| ctx_rh_posteriorcingulate_1 | 13.654428 | 2.499378e-03 | 2.344402e-04 |
| ctx_lh_caudalanteriorcingulate_1 | 13.624671 | 2.499378e-03 | 2.381602e-04 |
| ctx_lh_parstriangularis_1 | 13.601467 | 2.499378e-03 | 2.411680e-04 |
| ctx_lh_precentral_1 | 12.604922 | 3.676939e-03 | 4.068653e-04 |
| ctx_lh_superiorparietal_3 | 12.547910 | 3.676939e-03 | 4.193001e-04 |
| ctx_lh_precuneus_2 | 11.935809 | 4.716672e-03 | 5.792404e-04 |
| ctx_lh_posteriorcingulate_1 | 11.032925 | 7.106567e-03 | 9.350746e-04 |
| ctx_lh_frontalpole_1 | 10.619003 | 8.308002e-03 | 1.166035e-03 |
| ctx_lh_inferiortemporal_1 | 10.469119 | 8.470467e-03 | 1.263140e-03 |
| ctx_rh_superiorparietal_2 | 10.326392 | 8.633392e-03 | 1.363167e-03 |
| ctx_lh_precentral_4 | 9.177773 | 1.516415e-02 | 2.527358e-03 |
| ctx_lh_inferiorparietal_2 | 8.841523 | 1.661382e-02 | 3.031536e-03 |
| ctx_rh_frontalpole_1 | 8.824410 | 1.661382e-02 | 3.060440e-03 |
| ctx_lh_lateralorbitofrontal_1 | 8.444210 | 1.907932e-02 | 3.762134e-03 |
| ctx_rh_inferiortemporal_1 | 8.402166 | 1.907932e-02 | 3.849336e-03 |
| ctx_rh_inferiortemporal_2 | 8.187293 | 2.055647e-02 | 4.327679e-03 |
| ctx_rh_medialorbitofrontal_1 | 7.555923 | 2.683220e-02 | 6.114779e-03 |
| ctx_lh_supramarginal_2 | 7.554334 | 2.683220e-02 | 6.119625e-03 |
| ctx_lh_parsorbitalis_1 | 7.251578 | 3.001525e-02 | 7.231163e-03 |
| ctx_lh_superiortemporal_2 | 7.216228 | 3.001525e-02 | 7.372166e-03 |
| ctx_lh_rostralmiddlefrontal_2 | 6.961022 | 3.337984e-02 | 8.491363e-03 |
| ctx_rh_postcentral_1 | 6.835145 | 3.459521e-02 | 9.104002e-03 |
| ctx_rh_superiorfrontal_1 | 6.435582 | 4.182231e-02 | 1.137273e-02 |
| ctx_rh_superiorfrontal_3 | 6.175495 | 4.686116e-02 | 1.315401e-02 |
| ctx_rh_supramarginal_2 | 6.096944 | 4.748911e-02 | 1.374685e-02 |

Supplementary Table S4. Regional modal controllability association with chronological age in MDD patients.

| DV | F | p_cor | p |
|---|---|---|---|
| ctx_rh_inferiorparietal_2 | 37.325569 | 1.909615e-07 | 1.675101e-09 |
| ctx_rh_caudalanteriorcingulate_1 | 28.010158 | 9.267721e-06 | 1.625916e-07 |
| ctx_rh_rostralmiddlefrontal_2 | 24.699890 | 3.223849e-05 | 8.483813e-07 |
| ctx_rh_paracentral_1 | 22.480421 | 7.363334e-05 | 2.583626e-06 |
| ctx_lh_rostralmiddlefrontal_3 | 21.449470 | 9.925457e-05 | 4.353271e-06 |
| ctx_rh_superiorparietal_2 | 20.003847 | 1.719525e-04 | 9.050129e-06 |
| ctx_lh_paracentral_1 | 19.087578 | 2.347886e-04 | 1.441684e-05 |
| ctx_lh_precuneus_2 | 17.430661 | 4.796616e-04 | 3.366046e-05 |
| ctx_rh_fusiform_1 | 16.857291 | 5.728120e-04 | 4.522200e-05 |
| ctx_rh_medialorbitofrontal_1 | 15.531192 | 1.020696e-03 | 8.953470e-05 |
| ctx_lh_caudalanteriorcingulate_1 | 14.110583 | 1.778032e-03 | 1.870359e-04 |
| ctx_rh_superiortemporal_1 | 14.109750 | 1.778032e-03 | 1.871613e-04 |
| ctx_lh_precentral_4 | 13.939611 | 1.792982e-03 | 2.044628e-04 |
| ctx_lh_superiorfrontal_1 | 11.921089 | 4.797406e-03 | 5.891552e-04 |
| ctx_lh_precentral_1 | 11.741753 | 4.922560e-03 | 6.477053e-04 |
| ctx_rh_rostralanteriorcingulate_1 | 11.570555 | 5.051473e-03 | 7.089787e-04 |
| ctx_rh_parstriangularis_1 | 10.820303 | 7.077848e-03 | 1.055469e-03 |
| ctx_rh_frontalpole_1 | 10.067371 | 9.602155e-03 | 1.577449e-03 |
| ctx_lh_posteriorcingulate_1 | 10.039339 | 9.602155e-03 | 1.600359e-03 |
| ctx_lh_insula_2 | 9.736101 | 1.073597e-02 | 1.883503e-03 |
| ctx_rh_superiorparietal_3 | 9.076787 | 1.456973e-02 | 2.683897e-03 |
| ctx_rh_superiortemporal_2 | 8.264725 | 2.159615e-02 | 4.167678e-03 |
| ctx_lh_supramarginal_1 | 7.571802 | 3.016701e-02 | 6.086326e-03 |
| ctx_lh_lingual_2 | 7.434614 | 3.026175e-02 | 6.563486e-03 |
| ctx_lh_rostralanteriorcingulate_1 | 7.414124 | 3.026175e-02 | 6.636349e-03 |
| ctx_lh_inferiortemporal_1 | 7.240743 | 3.170041e-02 | 7.300815e-03 |
| ctx_rh_parsorbitalis_1 | 7.189977 | 3.170041e-02 | 7.507991e-03 |
| ctx_rh_lateralorbitofrontal_2 | 7.060497 | 3.193893e-02 | 8.065215e-03 |
| ctx_rh_precuneus_2 | 7.046926 | 3.193893e-02 | 8.124815e-03 |
| ctx_lh_inferiorparietal_2 | 6.567027 | 3.952053e-02 | 1.060176e-02 |
| ctx_lh_parstriangularis_1 | 6.542663 | 3.952053e-02 | 1.074681e-02 |
| ctx_rh_inferiorparietal_1 | 6.345402 | 4.273625e-02 | 1.199614e-02 |
| ctx_lh_superiortemporal_2 | 6.058722 | 4.865097e-02 | 1.408318e-02 |

Supplementary Table S5. Regional average controllability association with gender in healthy controls.

| DV | F | p_cor | p |
|---|---|---|---|
| ctx_rh_precentral_1 | 29.443157 | 0.000005 | 7.599849e-08 |
| ctx_lh_insula_1 | 29.003714 | 0.000005 | 9.459793e-08 |
| ctx_rh_parsopercularis_1 | 21.103925 | 0.000192 | 5.039476e-06 |
| ctx_rh_postcentral_1 | 19.952573 | 0.000258 | 9.060701e-06 |
| ctx_lh_paracentral_1 | 15.083175 | 0.002536 | 1.112220e-04 |
| ctx_lh_caudalmiddlefrontal_1 | 12.060457 | 0.010302 | 5.421964e-04 |
| ctx_lh_lingual_1 | 11.153631 | 0.014280 | 8.768306e-04 |
| ctx_lh_transversetemporal_1 | 10.898976 | 0.014308 | 1.004087e-03 |
| ctx_rh_paracentral_1 | 10.480137 | 0.015903 | 1.255482e-03 |
| ctx_rh_parsorbitalis_1 | 9.256815 | 0.027606 | 2.421567e-03 |
| ctx_lh_pericalcarine_1 | 8.600614 | 0.035804 | 3.454755e-03 |
| ctx_rh_rostralanteriorcingulate_1 | 8.092427 | 0.043288 | 4.556585e-03 |

Supplementary Table S6. Regional average controllability association with gender in MDD patients.

| DV | F | p_cor | p |
|---|---|---|---|
| ctx_rh_superiorparietal_2 | 28.118446 | 0.000011 | 1.540171e-07 |
| ctx_lh_paracentral_1 | 27.588644 | 0.000011 | 2.003768e-07 |
| ctx_rh_posteriorcingulate_1 | 25.908652 | 0.000018 | 4.626429e-07 |
| ctx_rh_caudalanteriorcingulate_1 | 18.397948 | 0.000511 | 2.049504e-05 |
| ctx_rh_precentral_1 | 18.221229 | 0.000511 | 2.243288e-05 |
| ctx_rh_postcentral_1 | 17.409728 | 0.000646 | 3.399421e-05 |
| ctx_rh_paracentral_1 | 15.974343 | 0.001159 | 7.115122e-05 |
| ctx_lh_precentral_4 | 15.092445 | 0.001600 | 1.122739e-04 |
| ctx_lh_posteriorcingulate_1 | 11.805766 | 0.007930 | 6.260193e-04 |
| ctx_lh_insula_1 | 9.902172 | 0.019633 | 1.722172e-03 |
| ctx_rh_lateralorbitofrontal_1 | 9.687542 | 0.020024 | 1.932170e-03 |
| ctx_rh_parsopercularis_1 | 9.138953 | 0.024656 | 2.595385e-03 |
| ctx_rh_supramarginal_2 | 7.879646 | 0.045084 | 5.141153e-03 |

Supplementary Table S7. Regional modal controllability association with gender in healthy controls.

| DV | F | p_cor | p |
|---|---|---|---|
| ctx_lh_insula_1 | 54.762759 | 3.866990e-11 | 3.392096e-13 |
| ctx_lh_paracentral_1 | 22.972955 | 1.113128e-04 | 1.952856e-06 |
| ctx_rh_paracentral_1 | 20.209989 | 3.019277e-04 | 7.945467e-06 |
| ctx_rh_precentral_1 | 19.050239 | 4.095830e-04 | 1.437133e-05 |
| ctx_lh_lingual_1 | 15.976227 | 1.595225e-03 | 6.996601e-05 |
| ctx_lh_transversetemporal_1 | 13.875475 | 3.967374e-03 | 2.088092e-04 |
| ctx_lh_superiortemporal_1 | 12.574513 | 6.732183e-03 | 4.133797e-04 |
| ctx_rh_rostralanteriorcingulate_1 | 11.930133 | 7.599321e-03 | 5.808747e-04 |
| ctx_lh_rostralmiddlefrontal_3 | 11.869071 | 7.599321e-03 | 5.999464e-04 |
| ctx_lh_pericalcarine_1 | 10.672392 | 1.291621e-02 | 1.133001e-03 |
| ctx_rh_posteriorcingulate_1 | 9.648390 | 2.032215e-02 | 1.960909e-03 |
| ctx_rh_postcentral_1 | 8.964606 | 2.694131e-02 | 2.835927e-03 |
| ctx_rh_parsorbitalis_1 | 8.418962 | 3.344086e-02 | 3.813432e-03 |
| ctx_rh_rostralmiddlefrontal_2 | 7.837538 | 4.265508e-02 | 5.238343e-03 |
| ctx_lh_postcentral_1 | 7.589240 | 4.562172e-02 | 6.002857e-03 |
| ctx_rh_transversetemporal_1 | 7.339845 | 4.906275e-02 | 6.886001e-03 |

Supplementary Table S8. Regional modal controllability association with gender in MDD patients.

| DV | F | p_cor | p |
|---|---|---|---|
| ctx_lh_paracentral_1 | 29.812114 | 0.000006 | 6.656394e-08 |
| ctx_rh_superiorparietal_2 | 28.811229 | 0.000006 | 1.092349e-07 |
| ctx_rh_posteriorcingulate_1 | 27.243799 | 0.000007 | 2.378552e-07 |
| ctx_lh_insula_1 | 27.158759 | 0.000007 | 2.481339e-07 |
| ctx_rh_caudalanteriorcingulate_1 | 20.767169 | 0.000140 | 6.138177e-06 |
| ctx_rh_paracentral_1 | 18.996339 | 0.000287 | 1.510021e-05 |
| ctx_lh_precentral_4 | 15.750004 | 0.001301 | 7.989105e-05 |
| ctx_rh_inferiorparietal_2 | 12.026760 | 0.007939 | 5.571246e-04 |
| ctx_rh_precentral_1 | 10.978378 | 0.011250 | 9.701581e-04 |
| ctx_lh_posteriorcingulate_1 | 10.946314 | 0.011250 | 9.868205e-04 |
| ctx_rh_rostralmiddlefrontal_2 | 10.478925 | 0.013115 | 1.265451e-03 |
| ctx_rh_lingual_2 | 9.199576 | 0.023863 | 2.511929e-03 |
| ctx_rh_postcentral_1 | 8.738560 | 0.028257 | 3.222264e-03 |
| ctx_lh_postcentral_1 | 8.348848 | 0.032416 | 3.980929e-03 |
| ctx_lh_precuneus_1 | 8.002895 | 0.035323 | 4.806458e-03 |
| ctx_rh_transversetemporal_1 | 7.946171 | 0.035323 | 4.957643e-03 |
| ctx_lh_transversetemporal_1 | 7.393712 | 0.045000 | 6.710519e-03 |
| ctx_rh_postcentral_2 | 7.103480 | 0.049870 | 7.874177e-03 |

Supplementary Table S9. Regional average controllability association with site in healthy controls.

| DV | F | p_cor | p |
|---|---|---|---|
| ctx_rh_precentral_1 | 29.443157 | 0.000005 | 7.599849e-08 |
| ctx_lh_insula_1 | 29.003714 | 0.000005 | 9.459793e-08 |
| ctx_rh_parsopercularis_1 | 21.103925 | 0.000192 | 5.039476e-06 |
| ctx_rh_postcentral_1 | 19.952573 | 0.000258 | 9.060701e-06 |
| ctx_lh_paracentral_1 | 15.083175 | 0.002536 | 1.112220e-04 |
| ctx_lh_caudalmiddlefrontal_1 | 12.060457 | 0.010302 | 5.421964e-04 |
| ctx_lh_lingual_1 | 11.153631 | 0.014280 | 8.768306e-04 |
| ctx_lh_transversetemporal_1 | 10.898976 | 0.014308 | 1.004087e-03 |
| ctx_rh_paracentral_1 | 10.480137 | 0.015903 | 1.255482e-03 |
| ctx_rh_parsorbitalis_1 | 9.256815 | 0.027606 | 2.421567e-03 |
| ctx_lh_pericalcarine_1 | 8.600614 | 0.035804 | 3.454755e-03 |
| ctx_rh_rostralanteriorcingulate_1 | 8.092427 | 0.043288 | 4.556585e-03 |

Supplementary Table S10. Regional average controllability association with site in MDD patients.

| DV | F | p_cor | p |
|---|---|---|---|
| ctx_rh_superiorparietal_2 | 28.118446 | 0.000011 | 1.540171e-07 |
| ctx_lh_paracentral_1 | 27.588644 | 0.000011 | 2.003768e-07 |
| ctx_rh_posteriorcingulate_1 | 25.908652 | 0.000018 | 4.626429e-07 |
| ctx_rh_caudalanteriorcingulate_1 | 18.397948 | 0.000511 | 2.049504e-05 |
| ctx_rh_precentral_1 | 18.221229 | 0.000511 | 2.243288e-05 |
| ctx_rh_postcentral_1 | 17.409728 | 0.000646 | 3.399421e-05 |
| ctx_rh_paracentral_1 | 15.974343 | 0.001159 | 7.115122e-05 |
| ctx_lh_precentral_4 | 15.092445 | 0.001600 | 1.122739e-04 |
| ctx_lh_posteriorcingulate_1 | 11.805766 | 0.007930 | 6.260193e-04 |
| ctx_lh_insula_1 | 9.902172 | 0.019633 | 1.722172e-03 |
| ctx_rh_lateralorbitofrontal_1 | 9.687542 | 0.020024 | 1.932170e-03 |
| ctx_rh_parsopercularis_1 | 9.138953 | 0.024656 | 2.595385e-03 |
| ctx_rh_supramarginal_2 | 7.879646 | 0.045084 | 5.141153e-03 |

Supplementary Table S11. Regional modal controllability association with site in healthy controls.

| DV | F | p_cor | p |
|---|---|---|---|
| ctx_lh_insula_1 | 54.762759 | 3.866990e-11 | 3.392096e-13 |
| ctx_lh_paracentral_1 | 22.972955 | 1.113128e-04 | 1.952856e-06 |
| ctx_rh_paracentral_1 | 20.209989 | 3.019277e-04 | 7.945467e-06 |
| ctx_rh_precentral_1 | 19.050239 | 4.095830e-04 | 1.437133e-05 |
| ctx_lh_lingual_1 | 15.976227 | 1.595225e-03 | 6.996601e-05 |
| ctx_lh_transversetemporal_1 | 13.875475 | 3.967374e-03 | 2.088092e-04 |
| ctx_lh_superiortemporal_1 | 12.574513 | 6.732183e-03 | 4.133797e-04 |
| ctx_rh_rostralanteriorcingulate_1 | 11.930133 | 7.599321e-03 | 5.808747e-04 |
| ctx_lh_rostralmiddlefrontal_3 | 11.869071 | 7.599321e-03 | 5.999464e-04 |
| ctx_lh_pericalcarine_1 | 10.672392 | 1.291621e-02 | 1.133001e-03 |
| ctx_rh_posteriorcingulate_1 | 9.648390 | 2.032215e-02 | 1.960909e-03 |
| ctx_rh_postcentral_1 | 8.964606 | 2.694131e-02 | 2.835927e-03 |
| ctx_rh_parsorbitalis_1 | 8.418962 | 3.344086e-02 | 3.813432e-03 |
| ctx_rh_rostralmiddlefrontal_2 | 7.837538 | 4.265508e-02 | 5.238343e-03 |
| ctx_lh_postcentral_1 | 7.589240 | 4.562172e-02 | 6.002857e-03 |
| ctx_rh_transversetemporal_1 | 7.339845 | 4.906275e-02 | 6.886001e-03 |

Supplementary Table S12. Regional modal controllability association with site in MDD patients.

| DV | F | p_cor | p |
|---|---|---|---|
| ctx_lh_paracentral_1 | 29.812114 | 0.000006 | 6.656394e-08 |
| ctx_rh_superiorparietal_2 | 28.811229 | 0.000006 | 1.092349e-07 |
| ctx_rh_posteriorcingulate_1 | 27.243799 | 0.000007 | 2.378552e-07 |
| ctx_lh_insula_1 | 27.158759 | 0.000007 | 2.481339e-07 |
| ctx_rh_caudalanteriorcingulate_1 | 20.767169 | 0.000140 | 6.138177e-06 |
| ctx_rh_paracentral_1 | 18.996339 | 0.000287 | 1.510021e-05 |
| ctx_lh_precentral_4 | 15.750004 | 0.001301 | 7.989105e-05 |
| ctx_rh_inferiorparietal_2 | 12.026760 | 0.007939 | 5.571246e-04 |
| ctx_rh_precentral_1 | 10.978378 | 0.011250 | 9.701581e-04 |
| ctx_lh_posteriorcingulate_1 | 10.946314 | 0.011250 | 9.868205e-04 |
| ctx_rh_rostralmiddlefrontal_2 | 10.478925 | 0.013115 | 1.265451e-03 |
| ctx_rh_lingual_2 | 9.199576 | 0.023863 | 2.511929e-03 |
| ctx_rh_postcentral_1 | 8.738560 | 0.028257 | 3.222264e-03 |
| ctx_lh_postcentral_1 | 8.348848 | 0.032416 | 3.980929e-03 |
| ctx_lh_precuneus_1 | 8.002895 | 0.035323 | 4.806458e-03 |
| ctx_rh_transversetemporal_1 | 7.946171 | 0.035323 | 4.957643e-03 |
| ctx_lh_transversetemporal_1 | 7.393712 | 0.045000 | 6.710519e-03 |
| ctx_rh_postcentral_2 | 7.103480 | 0.049870 | 7.874177e-03 |

Supplementary Table S13. Regional average controllability association with body mass index in MDD patients.

| DV | F | p_cor | p |
|---|---|---|---|
| ctx_lh_superiorfrontal_4 | 15.581079 | 0.010016 | 0.000088 |
| ctx_lh_posteriorcingulate_1 | 12.947477 | 0.019682 | 0.000345 |
| ctx_rh_superiortemporal_2 | 12.146568 | 0.019973 | 0.000526 |
| ctx_lh_lingual_1 | 11.197675 | 0.024725 | 0.000868 |
| ctx_lh_caudalmiddlefrontal_1 | 10.636984 | 0.026626 | 0.001168 |
| ctx_lh_precentral_1 | 9.894510 | 0.032957 | 0.001735 |
| ctx_rh_insula_2 | 8.909576 | 0.045189 | 0.002946 |
| ctx_rh_isthmuscingulate_1 | 8.704633 | 0.045189 | 0.003291 |
| ctx_rh_paracentral_1 | 8.555064 | 0.045189 | 0.003568 |

Supplementary Table S14. Regional modal controllability association with body mass index in MDD patients.

| DV | F | p_cor | p |
|---|---|---|---|
| ctx_lh_posteriorcingulate_1 | 13.485294 | 0.019637 | 0.000260 |
| ctx_rh_superiortemporal_2 | 12.182400 | 0.019637 | 0.000515 |
| ctx_rh_lateralorbitofrontal_1 | 12.177974 | 0.019637 | 0.000517 |
| ctx_lh_superiorfrontal_4 | 11.088033 | 0.023781 | 0.000919 |
| ctx_rh_insula_2 | 10.848571 | 0.023781 | 0.001043 |
| ctx_rh_isthmuscingulate_1 | 10.196834 | 0.028043 | 0.001476 |

Supplementary Table S15. Regional average controllability association with Familial Risk for Bipolar Disorder in MDD patients.

| DV | F | p_cor | p |
|---|---|---|---|
| ctx_rh_supramarginal_2 | 16.638199 | 0.005386 | 0.000051 |
| ctx_rh_inferiorparietal_1 | 15.425771 | 0.005386 | 0.000094 |
| ctx_rh_precuneus_2 | 11.371272 | 0.029937 | 0.000788 |

Supplementary Table S16. Regional modal controllability association with Familial Risk for Bipolar Disorder in MDD patients.

| DV | F | p_cor | p |
|---|---|---|---|
| ctx_rh_supramarginal_2 | 17.885003 | 0.003039 | 0.000027 |